\documentclass[letterpaper,12pt]{JHEP3}
\usepackage{amsmath,amssymb}
\usepackage{latexsym}
\usepackage{epsfig}
\usepackage{cite}
\usepackage[english]{babel}

\newcommand{\bld}[1]{\boldsymbol #1}

\newcommand{\eq}{\begin{equation}}
\newcommand{\feq}{\end{equation}}
\newcommand{\eqn}{\begin{eqnarray}}
\newcommand{\feqn}{\end{eqnarray}}
\newcommand{\arr}{\begin{eqnarray*}}
\newcommand{\farr}{\end{eqnarray*}}

\newcommand{\lp}{\left(}
\newcommand{\rp}{\right)}
\font\mybb=msbm10 at 12pt
\def\bb#1{\hbox{\mybb#1}}

\def\bR {\bb{R}}

\def\bC {\bb{C}}

\def\ga{\gamma}

\def\ep{\epsilon}

\def\ph{\varphi}
\def\th{\theta}

\def\De{\Delta}

\title{Magnetic charges in the AdS$_4$ superalgebra osp$(4|2)$}

\author{Giuseppe Dibitetto$^a$ and Dietmar Klemm$^{bc}$ \\
$^a$ Centre for Theoretical Physics, \\
\hspace*{0.15cm} University of Groningen, \\
\hspace*{0.15cm} Nijenborgh 4, 9747 AG Groningen, Netherlands. \\
$^b$ Dipartimento di Fisica dell'Universit\`a di Milano, \\
\hspace*{0.15cm} Via Celoria 16, I-20133 Milano. \\
$^c$ INFN, Sezione di Milano, Via Celoria 16, I-20133 Milano. \\
}
\preprint{IFUM-957-FT}

\abstract{We discuss the issue of how to include magnetic charges in the AdS$_4$
superalgebra osp$(4|2)$. It is shown that the usual way of introducing a pseudoscalar
central charge on the right hand side of the basic anticommutator does not
work, because this breaks SO$(2,3)$ covariance. We propose a way out by promoting
the magnetic charge to a vector charge, which amounts to enlarge osp$(4|2)$ to
the superconformal algebra su$(2,2|1)$. The conditions for 1/4, 1/2 and 3/4 BPS states
are then analyzed. These states form the boundary of the convex cone associated with
the Jordan algebra of 4$\times$4 complex hermitian matrices. An In\"on\"u-Wigner
contraction of the constructed superalgebra yields a known extension of the Poincar\'e
superalgebra containing electric and magnetic 0-brane charges as well as string- and
space-filling 3-brane charges. As an example, we show how some supersymmetric
AdS$_4$ black holes fit into the classification scheme of BPS states.
}


\begin{document}

\section{Introduction}

It is well-known that in asymptotically AdS$_4$ spacetimes there are extremal supersymmetric
black holes carrying magnetic charges \cite{Caldarelli:1998hg,Cacciatori:2009iz}\footnote{The
earliest reference in this topic is \cite{Romans:1991nq}, but the BPS magnetic monopoles
found there have naked singularities, and are thus not really black holes.}.
In order to describe such objects we have to find an extension of osp$(4|2)$ that includes
additional generators corresponding to such charges. A first attempt to do this can be found
in \cite{Kostelecky:1995ei}, where an extra generator $V$ (representing the magnetic charge)
was added to osp$(4|2)$ by deforming the basic anticommutator in the following way:
\eq
\{Q^i_{\alpha},Q^j_{\beta}\} = \delta^{ij}\left((\gamma^aM_{a4} - \gamma^{ab}M_{ab})C\right)_{\alpha\beta}
+ i\epsilon^{ij}\left(C_{\alpha\beta} U + i(C\gamma^5)_{\alpha\beta}V\right)\ ,
\label{basic-anti}
\feq
where $U$ denotes the electric charge.
The problem with this is that one of the super-Jacobi identities fails to hold. Indeed one
finds\footnote{Capital latin indices $A,B,\ldots$ refer to SO$(2,3)$ tensors and $\Gamma_A$
are SO$(2,3)$ Dirac matrices, cf.~appendix \ref{conv}. $M_{AB}$ denote the SO$(2,3)$ generators,
that are split in \eqref{basic-anti} into $M_{ab}$ and $M_{a4}$, with $a=0,\ldots,3$.}
\begin{displaymath}
\{Q^i_{\alpha},[Q^j_{\beta},M_{AB}]\} + {\text{permutations}} = \frac12V\epsilon^{ij}\left((\Gamma_{AB}
\Gamma^4C^{-1})_{\beta\alpha} - (\Gamma_{AB}\Gamma^4C^{-1})_{\alpha\beta}\right)\ ,
\end{displaymath}
which is clearly nonvanishing. This should not be surprising, since the magnetic term $\gamma^5V$
in \eqref{basic-anti} breaks SO$(2,3)$ covariance. ($\gamma^5$ is essentially one of the
SO$(2,3)$ generators). In section \ref{constr} we will propose
a way out by promoting the magnetic charge to a vector charge, and show that this construction
amounts to enlarging osp$(4|2)$ to the superconformal algebra su$(2,2|1)$. In \ref{anal-BPS}
we shall analyze the conditions for BPS states of this algebra, and compare
the results with some known supersymmetric black hole solutions. In the following section, the
geometrical interpretation of these BPS states is discussed, and it is shown that they form the
boundary of the convex cone associated with the Jordan algebra $J_4^{\bC}$ of
4$\times$4 complex hermitian matrices. We conclude in \ref{fin-rem} with some final remarks.
Appendix \ref{conv} contains our notations and conventions,
while in appendix \ref{app-osp} we briefly review the superalgebra osp$(4|N)$.

Here we shall not try to include nut charges. A way to do this in the case of the $N=2$
Poincar\'e superalgebra in four dimensions was proposed
in \cite{Argurio:2008zt}\footnote{Cf.~also \cite{Hull:1997kt}.}. Since
there exist also several supersymmetric black hole solutions in AdS with nonvanishing nut
charges \cite{AlonsoAlberca:2000cs}, it would be very interesting to see how AdS
superalgebras take them into account. We hope to come back to this point in a future publication.

\section{Construction of the superalgebra}
\label{constr}

In order to restore SO$(2,3)$ covariance, let us modify the basic anticommutator
of osp$(4|2)$ according to
\eq
\{Q^i_{\alpha},Q^j_{\beta}\} = \frac 12\delta^{ij}\left(\Gamma^{AB}C^{-1}\right)_{\alpha\beta}
M_{AB} + (C^{-1})_{\alpha\beta}U\epsilon^{ij} + \epsilon^{ij}\left(\Gamma^AC^{-1}\right)_{\alpha\beta}V_A\ ,
\label{QiQj}
\feq
where $Q^i_{\alpha}$ ($i=1,2$) are Majorana spinors and $V_A$ represent now magnetic
vector charges. Using the representation of the SO$(2,3)$ gamma matrices $\Gamma^A$
in terms of SO$(1,3)$ Dirac matrices $\gamma^a$ given in appendix \ref{conv}, we get
\begin{displaymath}
\Gamma^AC^{-1}V_A = \gamma^5C^{-1}V_4 + \gamma^5\gamma^aC^{-1}V_a\ ,
\end{displaymath}
and hence the first term on the rhs corresponds to the extension \eqref{basic-anti} proposed
in \cite{Kostelecky:1995ei}, whereas the second one is necessary for SO$(2,3)$ covariance.
Provided that $V^2\equiv V^AV_A$ is negative, one can
always set $V_a=0$ by an SO$(2,3)$ transformation, so that in this case \eqref{basic-anti}
is just a sort of gauge-fixed version of \eqref{QiQj}. Of course, such a gauge fixing spoils
covariance under SO$(2,3)$, which is broken down to the Lorentz group SO$(1,3)$.

The remaining commutation relations can be fixed by imposing the super-Jacobi identities.
This leads to
\begin{eqnarray}
[M_{AB},M_{CD}] &=& \eta_{AC}M_{BD} + \eta_{BD}M_{AC} - \eta_{AD}M_{BC} - \eta_{BC}M_{AD}\ ,
\nonumber \\
{}[M_{AB},Q^i_{\alpha}] &=& \frac 12{(\Gamma_{AB})_{\alpha}}^{\beta}Q^i_{\beta}\ , \qquad
[U,Q^k_{\alpha}] = -\frac32\epsilon_{ij}\delta^{jk}Q^i_{\alpha}\ , \nonumber \\
{}[V_A,Q^i_{\alpha}] &=& \frac 12{(\Gamma_A)_{\alpha}}^{\beta}\delta^{ik}\epsilon_{kj}Q^j_{\beta}\ ,\qquad
[V_A,V_B] = M_{AB}\ , \nonumber \\
{}[M_{AB},V_C] &=& \eta_{AC}V_B - \eta_{BC}V_A\ , \qquad
[U,M_{AB}] = [U,V_A] = 0\ . \label{superalg}
\end{eqnarray}
This superalgebra is actually isomorphic to the superconformal algebra su$(2,2|1)$
in four dimensions. The latter, which has the generators $P_a$, $J_{ab}$, $D$, $K_a$,
$Q_{\alpha}$, $S_{\alpha}$ plus an internal U$(1)$ symmetry generator $A$,
is given by the Lorentz group together with \cite{West:1998ey}
\begin{eqnarray}
[J_{ab},P_c] &=& \eta_{bc}P_a - \eta_{ac}P_b\ , \qquad
[J_{ab},K_c] = \eta_{bc}K_a - \eta_{ac}K_b\ , \nonumber \\
{}[D,P_a] &=& -P_a\ , \qquad [D,K_a] = K_a\ , \nonumber \\
{}[P_a,K_b] &=& -2J_{ab} + 2\eta_{ab}D\ , \qquad
[K,K] = [P,P] = 0\ , \nonumber \\
{}\left[Q_{\alpha},J_{ab}\right] &=& \frac12{(\gamma_{ab})_{\alpha}}^{\beta}Q_{\beta}\ , \qquad
\left[S_{\alpha},J_{ab}\right] = \frac12{(\gamma_{ab})_{\alpha}}^{\beta}S_{\beta}\ , \nonumber \\
\{Q_{\alpha},Q_{\beta}\} &=& -2(\gamma^aC^{-1})_{\alpha\beta}P_a\ , \qquad
\{S_{\alpha},S_{\beta}\} = 2(\gamma^aC^{-1})_{\alpha\beta}K_a\ , \nonumber \\
{}[Q_{\alpha},D] &=& \frac12 Q_{\alpha}\ , \qquad [S_{\alpha},D] = -\frac12 S_{\alpha}\ , \nonumber \\
{}[Q_{\alpha},K_a] &=& -{(\gamma_a)_{\alpha}}^{\beta}S_{\beta}\ , \qquad
[S_{\alpha},P_a] = {(\gamma_a)_{\alpha}}^{\beta}Q_{\beta}\ , \nonumber \\
{}[Q_{\alpha},A] &=& -\frac{3i}4{(\gamma^5)_{\alpha}}^{\beta}Q_{\beta}\ , \qquad
[S_{\alpha},A] = \frac{3i}4{(\gamma^5)_{\alpha}}^{\beta}S_{\beta}\ , \nonumber \\
\{Q_{\alpha},S_{\beta}\} &=& -2(C^{-1})_{\alpha\beta}D + (\gamma^{ab}C^{-1})_{\alpha\beta}J_{ab}
+4i(\gamma^5C^{-1})_{\alpha\beta}A\ .
\end{eqnarray}
The isomorphism with \eqref{QiQj}, \eqref{superalg} is provided by
\begin{eqnarray}
M_{ab} &=& -J_{ab}\ , \qquad M_{a4} = -\frac 12(P_a - K_a)\ , \qquad V_a = -\frac 12(P_a + K_a)\ ,
\qquad V_4 = D\ , \nonumber \\
Q^1 &=& \frac12(Q - S)\ , \qquad Q^2 = -\frac12\gamma^5(Q + S)\ , \qquad U = 2iA\ . \label{iso}
\end{eqnarray}
It is interesting to perform an In\"on\"u-Wigner contraction of the superalgebra
\eqref{QiQj}, \eqref{superalg}. This is done by rescaling
\eq
Q^i_{\alpha} \to \lambda^{1/2}Q^i_{\alpha}\ , \qquad M_{a4} \to \lambda M_{a4}\ , \qquad
U \to \lambda U\ , \qquad V_A \to \lambda V_A\ ,
\feq
and then taking the limit $\lambda\to\infty$, yielding
\begin{eqnarray}
\{Q^i_{\alpha},Q^j_{\beta}\} &=& -\delta^{ij}(\gamma^aC^{-1})_{\alpha\beta}P_a + \epsilon^{ij}
(\gamma^5\gamma^aC^{-1})_{\alpha\beta}V_a + \epsilon^{ij}U(C^{-1})_{\alpha\beta} +
\epsilon^{ij}V_4(\gamma^5C^{-1})_{\alpha\beta}\ , \nonumber \\
{}[M_{ab},M_{cd}] &=& \eta_{ac}M_{bd} + \eta_{bd}M_{ac} - \eta_{ad}M_{bc} - \eta_{bc}M_{ad}\ ,
\nonumber \\
{}[M_{ab},P_c] &=& \eta_{ac}P_b - \eta_{bc}P_a\ , \qquad [M_{ab},V_c] = \eta_{ac}V_b - \eta_{bc}V_a\ ,
\nonumber \\
{}[M_{ab},Q^i_{\alpha}] &=& \frac12{(\gamma_{ab})_{\alpha}}^{\beta}Q^i_{\beta} \label{Inonu}
\end{eqnarray}
as the only nonvanishing (anti)commutation relations\footnote{Here we defined $P_a=-M_{a4}$.}.
\eqref{Inonu} is a subcase of the extended Poincar\'e superalgebra considered in \cite{Gauntlett:1999dt},
with the spatial components of $V_a$ representing string charges, whereas $V_0$ is a charge for
space-filling 3-branes \cite{Hull:1997kt}. In the AdS version \eqref{QiQj}, \eqref{superalg}, these
string- and brane charges are thus unified with the magnetic 0-brane charge in the vector $V_A$.
Note in this context that the inclusion of brane charges in AdS$_5$ superalgebras was discussed
in \cite{Ferrara:1999si,Lee:2004jx}.

\section{Analysis of BPS states}
\label{anal-BPS}

In our conventions, Majorana spinors are real. Hence, we can define a single complex
Dirac supercharge by
\eq
Q = \frac1{\sqrt2}(Q^1 + iQ^2)\ ,
\feq
and the only nontrivial anticommutator (cf.~\eqref{QiQj}) becomes
\begin{eqnarray}
\{Q_{\alpha},Q^{\star}_{\beta}\}&=&\frac12\lp\ga^{ab}C^{-1}\rp_{\alpha\beta}
M_{ab} - \lp\ga^aC^{-1}\rp_{\alpha\beta} P_a - i\lp C^{-1}\rp_{\alpha\beta}
U \nonumber \\
&& - i\lp\ga^5\ga^a C^{-1}\rp_{\alpha\beta} V_a - i\lp\ga^5C^{-1}\rp_{\alpha\beta} V_4\ ,
\label{QQ}
\end{eqnarray}
where we have set $M_{a4}=-P_a$. When there is a multiplet of BPS states, some
combinations of the supercharges have to be represented trivially, i.e., they
have to vanish. This means that the positive semi-definite
hermitian matrix $\{Q_{\alpha},Q^{\star}_{\beta}\}$ is not of maximal rank; the right
hand side of \eqref{QQ} must have at least one vanishing eigenvalue.

In order to compute the determinant of $\{Q,Q^{\star}\}$, we choose the explicit
representation for the gamma matrices given in appendix \ref{conv}, and define
\eq
J^i=-\frac12\ep^{ijk}M_{jk}\,,\qquad K^i=-M^{0i}\,,\qquad
H=P^0\,,\qquad V^A=(V^0,\bld{V},W)\ .
\feq
Note that $\det\{Q,Q^{\star}\}$ is manifestly $\text{SL}(4,\bC)$ invariant, but the subgroup
keeping $H$ fixed is its maximal compact $\text{SU}(4)$ subgroup. $\det\{Q,Q^{\star}\}$
is thus a fourth-order polynomial in $H$ with coefficients that are homogeneous
polynomials in the three algebraically-independent $\text{SU}(4)$ invariants that can
be constructed from $\bld{J}$, $\bld{P}$,
$\bld{K}$, $\bld{V}$, $U$, $V^0$ and $W$. In fact we find\footnote{The
absence of the $H^3$ term in \eqref{poly} is due to the fact that
it represents the characteristic polynomial of a traceless matrix. }
\eq
\det(\{Q,Q^{\star}\})=H^4-2aH^2+8bH+(a^2-4c)\ ,
\label{poly}
\feq
where
\begin{align}
a&=|\bld{J}|^2+|\bld{P}|^2+|\bld{K}|^2+|\bld{V}|^2+U^2+(V^0)^2+W^2\,, \label{a} \\
\notag\\
b&=\bld{P}\cdot(\bld{J}\times\bld{K})+\bld{V}\cdot(U\bld{J}+W\bld{K}+V^0\bld{P})\,,\\
\notag\\
c&=\lp
(V^0)^2+W^2+U^2\rp|\bld{V}|^2+|U\bld{J}+W\bld{K}+V^0\bld{P}|^2+
\notag\\
&+|\bld{V}\times\bld{P}|^2-2\bld{V}\cdot(U\bld{P}\times\bld{K}+V^0\bld{K}\times\bld{J}+W\bld{P}\times\bld{J})+
\notag\\
&+|\bld{K}\times\bld{P}|^2+|\bld{J}\times\bld{P}|^2+|\bld{K}\times\bld{J}|^2+|\bld{V}\times\bld{J}|^2+|\bld{V}\times\bld{K}|^2\,.
\end{align}
The positivity condition for the hermitian matrix $\{Q,Q^{\star}\}$
imposes a lower bound on $H$ in terms of the invariants $a,b,c$,
\eq
H\geq E(a,b,c)\ , \label{bound}
\feq
where $E$ is the largest root of \eqref{poly},
which has to be non-negative since the sum of the roots
vanishes. \eqref{bound} is easily seen by writing \eqref{QQ} in the form
\begin{displaymath}
\{Q_{\alpha},Q^{\star}_{\beta}\} = H\delta_{\alpha\beta} - \Lambda_{\alpha\beta}\ ,
\end{displaymath}
and going to a basis in which $\Lambda_{\alpha\beta}$ is diagonal.
Configurations saturating \eqref{bound} are BPS, the
number of supersymmetries preserved being equal to the multiplicity
of the eigenvalue $E$. Since the characteristic polynomial \eqref{poly}
is the same as the one for the centrally extended $N=1$ super-Poincar\'e
algebra \cite{Gauntlett:2000ch} (although the invariants $a,b,c$ are of course
different), and the conditions obeyed by $a,b,c$ in order to have 1/4, 1/2 or 3/4
supersymmetry were determined in \cite{Gauntlett:2000ch}, we shall not repeat
such an analysis here, but instead give explicit examples of black holes
saturating the bound \eqref{bound}.

\subsection{Supersymmetric AdS black holes in $N=2$ gauged supergravity}

We shall now compare the above results with some known
BPS black hole solutions in minimal $N=2$ gauged
supergravity \cite{Kostelecky:1995ei,Caldarelli:1998hg}.
We would like to stress that the present classification only
analyzes the supersymmetry algebra and thus it provides
model-independent constraints on BPS configurations but it is still
regardless of possible further constraints imposed by certain
realizations thereof.

\subsubsection{The Kerr-Newman-AdS$_4$ solution}

The metric and electromagnetic vector potential of the
Kerr-Newman-AdS$_4$ black hole are given respectively
by (cf.~e.g.~\cite{Caldarelli:1999xj})
\begin{displaymath}
\textrm{d}s^2=-\frac{\De_r}{\rho^2}\lp \textrm{d}t-\frac{a\sin^2
\th}{\Xi}\textrm{d}\ph\rp^2+\frac{\rho^2}{\De_r}\textrm{d}r^2
+\frac{\rho^2}{\De_\th}\textrm{d}\th^2+\frac{\De_\th\sin^2\th}{\rho^2}\lp
a\textrm{d}t-\frac{r^2+a^2}{\Xi}\textrm{d}\ph\rp^2\ ,
\end{displaymath}
\eq
A = -\frac{q_er}{\rho^2}\left(\textrm{d}t - \frac{a\sin^2\theta}{\Xi}\textrm{d}\varphi\right)
- \frac{q_m\cos\theta}{\rho^2}\left(a\textrm{d}t - \frac{r^2+a^2}{\Xi}\textrm{d}\varphi\right)\ ,
\feq
where
\begin{displaymath}
\rho^2 = r^2 + a^2\cos^2\th\ , \qquad \De_\th = 1-\frac{a^2}{l^2}\cos^2\th\ ,
\end{displaymath}
\begin{displaymath}
\De_r=\lp r^2+a^2\rp\lp 1+ \frac{r^2}{l^2}\rp - 2mr+q_{\text e}^2 + q_{\text m}^2\ , \qquad
\Xi=1-\frac{a^2}{l^2}\ .
\end{displaymath}
The parameters $m$, $a$, $q_{\text e}$ and $q_{\text m}$ are related to the
mass, angular momentum, electric and magnetic charges respectively\footnote{We
apologize for using the same symbol $a$ for the angular momentum parameter
and the invariant \eqref{a}, but the meaning should be clear from the context.}
(see below). This solution admits Killing spinors if \cite{Caldarelli:1998hg}
\eq
q_{\text m} = 0\ , \qquad m^2 = \left(1 + \frac al\right)^2q_{\text e}^2\ .
\label{cond-kill}
\feq
Plugging this into the extremality condition
\begin{eqnarray}
m_{\text{extr.}} &=& \frac l{3\sqrt 6}\left(\sqrt{\left(1 + \frac{a^2}{l^2}\right)^2 + \frac{12}{l^2}
(a^2 + q_{\text e}^2 + q_{\text m}^2)} + \frac{2a^2}{l^2} + 2\right) \nonumber \\
&&\cdot\left(\sqrt{\left(1 + \frac{a^2}{l^2}\right)^2 + \frac{12}{l^2}(a^2 + q_{\text e}^2 +
q_{\text m}^2)} - \frac{a^2}{l^2} - 1\right)^{1/2} \nonumber
\end{eqnarray}
yields
\eq
m^2 = al\left(1 + \frac al\right)^4\ . \label{bps-extr}
\feq
The mass, angular momentum and electric charge of the Kerr-Newman-AdS$_4$
black hole are given by \cite{Caldarelli:1999xj}
\eq
M=\frac m{\Xi^2}\ , \qquad J=\frac{am}{\Xi^2}\ , \qquad Q_{\text e}=\frac{q_e}{\Xi}\,.
\feq
Using \eqref{cond-kill} and \eqref{bps-extr} gives
\eq
M = \frac{\sqrt{al}}{\left(1 - \frac al\right)^2}\ , \qquad
J = \frac{a\sqrt{al}}{\left(1 - \frac al\right)^2}\ , \qquad
Q_{\text e} = \frac{\sqrt{al}}{\left(1 - \frac al\right)}\ ,
\feq
and thus
\eq
M = Q_{\text e} + \frac Jl\,. \label{KN-BPS}
\feq
To check the consistency of this relation with the constraints coming from
the superalgebra, we have to go back to \eqref{poly} where everything is
vanishing except $H=M$, $|\bld{J}|=J$ and $U=Q_{\text e}$. In this case,
\eqref{poly} boils down to
\eq
\det\{Q,Q^{\star}\} = M^4 - 2(J^2+Q_{\text e}^2)M^2 + (J^2-Q_{\text e}^2)^2\ ,
\feq
whose largest root is $M=Q_{\text e}+J$, which exactly coincides with
\eqref{KN-BPS} after setting $l=1$ for the AdS curvature radius (which is the
choice made in the superalgebra \eqref{QiQj}, \eqref{superalg}). This solution
preserves one quarter of the supersymmetry.

\subsubsection{Static magnetic topological black holes}

Let us now consider magnetically charged BPS black holes in minimal
$N=2$, $D=4$ gauged supergravity\footnote{Generalizations to the case
of gauged supergravity with matter coupling can be found in \cite{Cacciatori:2009iz}.}.
These belong to a class of solutions known as topological
black holes \cite{Vanzo:1997gw},
whose horizon can be an arbitrary Riemann surface of constant
curvature $\kappa$. In what follows we shall be interested in the
case $\kappa<0$. Then the metric and the electromagnetic gauge potential
read respectively (cf.~e.g.~\cite{Caldarelli:1998hg})
\eq
\textrm{d}s^2=-V(r)\textrm{d}t^2+\frac{\textrm{d}r^2}{V(r)}+
r^2(\textrm{d}\theta^2+\sinh^2\theta\textrm{d}\varphi^2)\ , \label{magn-BH}
\feq
\begin{displaymath}
A = -\frac{q_{\text e}}r\textrm{d}t + q_{\text m}\cosh\theta\textrm{d}\varphi\ ,
\end{displaymath}
where
\eq
V(r)=-1-\frac{2m}r+\frac{q_{\text e}^2+q_{\text m}^2}{r^2}+\frac{r^2}{l^2}\ .
\feq
If the horizon is compactified to a Riemann surface of genus $g>1$, the
black hole mass, electric and magnetic charges are given by
\eq
M=m(g-1)\ , \qquad Q_{\text e}=q_{\text e}(g-1)\ , \qquad
Q_{\text m}=q_{\text m}(g-1)\ . \label{charges-top}
\feq
The Killing spinor equations for the above geometry were solved
in \cite{Caldarelli:1998hg}, and it was found that they imply the constraints
\eq
m = q_{\text e} = 0\ , \qquad q_{\text m} = \pm\frac l2 \label{constr-magn}
\feq
on the parameters. If \eqref{constr-magn} is satisfied, one has a 1/4 BPS,
extremal massless black hole carrying only magnetic charge,
with an event horizon at $r=l/\sqrt2$. This is a solitonic
object, since it does not admit a limit where the cosmological
constant $\Lambda=-3/l^2$ goes to zero. Notice that the mass
appearing in \eqref{charges-top} is not to be identified with the
Hamiltonian $H=M_{04}$. To see this, consider the asymptotic form
of \eqref{magn-BH} for $r\to\infty$,
\eq
\textrm{d}s^2=-\left(-1+\frac{r^2}{l^2}\right)\textrm{d}t^2+\frac{\textrm{d}r^2}
{-1+\frac{r^2}{l^2}}+r^2(\textrm{d}\theta^2+\sinh^2\theta\textrm{d}\varphi^2)\ ,
\feq
which is obtained as induced metric on the hypersurface $\eta_{AB}X^AX^B=-l^2$
in $\bR^5_2$ by choosing the parametrization
\begin{eqnarray}
X^0 &=& r\cosh\theta\ , \qquad X^1 = r\sinh\theta\cos\varphi\ , \qquad X^2 =
r\sinh\theta\sin\varphi\ , \nonumber \\
X^3 &=& \sqrt{r^2-l^2}\cosh\frac tl\ , \qquad X^4 = \sqrt{r^2-l^2}\sinh\frac tl\ .
\label{param-hyp}
\end{eqnarray}
On the other hand, setting
\begin{eqnarray}
X^0 &=& \sqrt{\rho^2+l^2}\sin\frac{\tau}l\ , \qquad X^4 = \sqrt{\rho^2+l^2}\cos\frac{\tau}l\ , 
\nonumber \\
X^1 &=& \rho\sin\vartheta\cos\phi\ , \qquad X^2 = \rho\sin\vartheta\sin\phi\ , \qquad
X^3 = \rho\cos\vartheta\ ,
\end{eqnarray}
one gets AdS$_4$ in global coordinates,
\eq
\textrm{d}s^2=-\left(1+\frac{\rho^2}{l^2}\right)\textrm{d}\tau^2+\frac{\textrm{d}\rho^2}
{1+\frac{\rho^2}{l^2}}+\rho^2(\textrm{d}\vartheta^2+\sin^2\vartheta\textrm{d}\phi^2)\ .
\feq
The AdS Killing vectors are
\begin{displaymath}
\xi_{AB} = X_A\partial_B - X_B\partial_A\ .
\end{displaymath}
In the parametrization \eqref{param-hyp} we have $\xi_{34}=l\partial_t$, and hence
the mass \eqref{charges-top}, which was computed with respect to $\partial_t$,
corresponds to the generator $M_{34}=-P_3$. The Hamiltonian $H=M_{04}$, instead,
generates translations in the global time $\tau$, since $\xi_{04}=l\partial_{\tau}$. In the coordinates
\eqref{param-hyp}, one has
\eq
\xi_{04} = \sqrt{r^2-l^2}\sinh\frac tl\left(-\frac{\sinh\theta}r\partial_{\theta} + \cosh\theta
\partial_r\right) - \frac{r\cosh\theta}{\sqrt{\frac{r^2}{l^2}-1}}\cosh\frac tl\partial_t\ .
\label{xi04}
\feq
Setting everything to zero in \eqref{poly} except the magnetic charge $W$ and
possible boost charges $\bld{K}$, we find that the largest root is given by
\eq
H = |\bld{K}| + |W|\ . \label{HKW}
\feq
This is of course strictly positive, so there should be a nonvanishing conserved charge
associated to \eqref{xi04}. Since the holographic stress tensor of the solution
\eqref{magn-BH} vanishes for $m=0$ \cite{Emparan:1999gf}\footnote{Actually,
\cite{Emparan:1999gf} considers only uncharged topological black holes. However,
since the charges $q_{\text e}, q_{\text m}$ give only subleading contributions to
the metric compared with the mass term, they do not alter the holographic stress tensor.},
it is difficult to see how such a nonzero value of $H$ could arise. One possibility is
that there exist finite counterterms that can be added to the usual counterterms used
in the holographic renormalization procedure (cf.~e.g.~\cite{Balasubramanian:1999re}),
which would then contribute to the stress tensor and thus to $H$.

Apart from this issue, it is also clear from \eqref{HKW} that we need $\bld{K}\neq0$
(alternatively one could have $\bld{P}\neq 0$), since otherwise \eqref{HKW}
becomes a double root, but we know from the Killing spinor equations
that the above black hole preserves only one quarter of the
supersymmetry \cite{Caldarelli:1998hg}. This seems to suggest that
the solution \eqref{magn-BH} indeed carries boost charges as well.
This is not incongruous, since in the description in terms of Poincar\'e
coordinates, $t$ in \eqref{magn-BH} becomes Rindler time \cite{Emparan:1999gf},
and a translation in Rindler time corresponds to a boost in Minkowski time.
We shall leave a deeper investigation of these points for a future publication.

Notice finally that \eqref{constr-magn} gives a Dirac-type quantization
condition for the magnetic charge, which comes from the minimal
coupling of the gravitino to the gauge field (with coupling constant
$1/l$), and does not seem to have any model-independent origin.

\section{Geometrical interpretation of BPS states}

In this section we shall discuss the geometry associated with BPS representations
of the algebra \eqref{QiQj}, \eqref{superalg}. To this end, we need the concepts of convex cones
and Jordan algebras, that will be introduced below. Our presentation follows closely
ref.~\cite{Gauntlett:2000ch}, to which we refer for more details.

\subsection{Convex cones}

{\bf Definition} - Let $V$ be a real vector space. A convex cone ${\cal C}\subset V$ is
an $n$-dimensional subspace such that
\begin{displaymath}
\text{i})\quad x\in {\cal C}\ , \,\lambda\in\bR^+\Rightarrow\lambda x\in{\cal C}\ , \qquad
\text{ii})\quad x,y\in {\cal C}\Rightarrow x+y\in{\cal C}\ .
\end{displaymath}
If $\langle\,,\rangle:V\times V^{\ast}\rightarrow\bR$ is a bilinear map, one can
define the dual cone ${\cal C}^{\ast}$ by
\eq
{\cal C}^{\ast} := \{y\in V^{\ast}|\langle x,y\rangle>0\,\,\forall x\in {\cal C}\}\ .
\feq
Once $V^{\ast}$ is equipped with a translation-invariant measure $d^ny$, it becomes possible
to define a characteristic function $\omega$ associated with the convex cone in the
following way:
\eq
\omega^{-1}(x) := \int_{{\cal C}^{\ast}}e^{-\langle x,y\rangle}d^ny\ . \label{omega}
\feq
Since the map $D:x\mapsto\lambda x$, $\lambda\in\bR^+$, is an automorphism of
$\cal C$, $\omega$ is a homogeneous function of degree $n$. The immediate proof
of this statement consists in a simple change of variables in the integral \eqref{omega}.
A consequence of this observation is $\langle x,\pi(x)\rangle=1\,\,\forall x\in\cal C$, where
\eq
\pi(x) := \frac 1n\frac{\partial\log\omega}{\partial x}\ .
\feq
Thus, $\pi\in{\cal C}^{\ast}$, and by letting $x$ range over all vectors in $\cal C$, $\pi(x)$
spans the whole ${\cal C}^{\ast}$. One can now introduce a metric on $\cal C$ by
\eq
g_{ij} = -\frac1n\partial_i\partial_j\log\omega(x)\ .
\feq
It is straightforward to shew that
\eq
\pi_j = x^ig_{ij}\ .
\feq
A simple example of a convex cone is the forward light cone in four-dimensional
Minkowski spacetime, where $V\cong\bR^4$ and $\omega={\cal N}^2$, where
${\cal N}(x)=-\eta_{\mu\nu}x^{\mu}x^{\nu}$. Note that, in general, $\cal C$ is foliated
by hypersurfaces of constant $\omega$, with the limiting case $\omega=0$ representing
the boundary of the cone. If the hypersurfaces of constant $\omega$ are homogeneous
spaces, the convex cone is called homogeneous.

\subsection{Jordan algebras}

{\bf Definition} - A Jordan algebra of dimension $n$ and degree $\nu$ is a triplet
$(J,\circ,{\cal N})$ such that\\
i) $J$ is an $n$-dimensional real vector space,\\
ii) $\circ$ is a commutative, power associative and bilinear product,\\
iii) $\cal N$ is a norm with the property of being a homogeneous polynomial of degree $\nu$.\\

There are four infinite series of simple Jordan algebras, realizable as matrices with the Jordan
product being the anticommutator: The degree 2 algebras $\Sigma(n)$ and the series
$J_k^{\bR}$, $J_k^{\bC}$ and $J_k^{\bb{H}}$; in addition there is also one exceptional Jordan
algebra $J_3^{\bb{O}}$. $\Sigma(n)$ is an $n$-dimensional algebra spanned by
$(1,\sigma_1,\ldots,\sigma_{n-1})$, where $\sigma_a$ are sigma matrices
of an $n$-dimensional Minkowski spacetime. $J_k^{\bb{K}}$ instead are realized by
hermitian $k\times k$ matrices over the field $\bb{K}$, with norm given by the determinant.

With any Jordan algebra we can associate a selfdual homogeneous convex cone
${\cal C}(J)$ defined (just like in the case of Lie groups and Lie algebras) by an
exponential mapping,
\eq
\exp: J \rightarrow {\cal C}(J)\ , \qquad x \mapsto \exp(x) := \sum_{k=0}^{\infty}\frac{x^k}{k!}\ ,
\feq
where
\begin{displaymath}
x^{k+1} = x^k\circ x\ , \qquad x^0 = 1_J\ .
\end{displaymath}
In this particular case, the characteristic function is
\eq
\omega = {\cal N}^{n/\nu}\ ,
\feq
and thus the boundary of the cone is made by the elements of $J$ with vanishing norm.
The cone is foliated by copies of the homogeneous space $\text{Str}(J)/\text{Aut}(J)$, where
$\text{Str}(J)$ denotes the invariance group of the norm $\cal N$ (the structure group of the
algebra), and $\text{Aut}(J)$ is the automorphism group, i.e., the subgroup of
$\text{Str}(J)$ leaving the identity element $1_J$ invariant.
In addition to $\text{Aut}(J)$ and $\text{Str}(J)$, there is a third group naturally
associated to a Jordan algebra, which is the M\"obius group $\text{Mo}(J)$ of
fractional linear transformations,
\eq
J\ni X \mapsto X' = (AX+B)(CX+D)^{-1}\ . \label{moebius}
\feq
One has therefore the sequence of groups
\eq
\text{Aut}(J) \subset \text{Str}(J) \subset \text{Mo}(J)\ , \label{sequence}
\feq
that can be interpreted as generalized rotation, Lorentz and conformal groups
respectively \cite{Gunaydin:1992zh}.

As an example motivating this, let us again consider
the forward light cone in four-dimensional Minkowski spacetime. This may be
interpreted as the selfdual homogeneous convex cone associated
with the Jordan algebra $J_2^{\bC}$ of hermitian $2\times2$ matrices over $\bC$ by means
of the identification
\eq
X = (X^0,\bld{X}) \quad\leftrightarrow\quad X^{\mu}\sigma_{\mu} = X^01 + X^i\sigma^i\ .
\feq
Since in this case the norm is given by the determinant, the structure group
is SL$(2,\bC)$, which acts on $2\times2$ matrices by conjugation. The automorphism group
fixing the identity matrix is then SU$(2)$, so that the cone is foliated by copies of
SL$(2,\bC)$/SU$(2)$. In order to obtain the M\"obius group, one imposes hermiticity of
$X'$ in \eqref{moebius}, which requires that
\eq
\left(\begin{array}{cc} A & B \\ C & D\end{array}\right) \in \text{SU}(2,2)\ .
\feq
This is the conformal group acting on the compactification of Minkowski space.
The sequence \eqref{sequence} becomes here
\eq
\text{SU}(2) \subset \text{SL}(2,\bC) \subset \text{SU}(2,2)\ .
\feq

\subsection{The geometry of BPS states}

The basic anticommutator \eqref{QQ} can be rewritten in terms of a
hermitian bispinor $Z$ as $\{Q_{\alpha},Q^{\star}_{\beta}\}=Z_{\alpha\beta}$.
Since $\{Q_{\alpha},Q^{\star}_{\beta}\}$ is
positive semi-definite, $Z$ is a vector in a convex cone, whose boundary
corresponds to the BPS condition $\det Z=0$.
To see this, consider the Jordan algebra $J_4^{\bC}$ of hermitian $4\times4$ matrices.
Its associated convex cone $\cal C$ consists of those complex hermitian matrices that
can be written as exponentials of other hermitian matrices, or, in other words, that have
only nonnegative eigenvalues. One has thus $Z\in{\cal C}$.

The structure group of $J_4^{\bC}$ is SL$(4,\bC)$, whereas $\text{Aut}(J_4^{\bC})\cong\text{SU}(4)$,
and thus the cone is foliated by copies of the symmetric space SL$(4,\bC)$/SU$(4)$.
In order to obtain the M\"obius group, we require hermiticity of $X'$ in \eqref{moebius}.
This gives the conditions
\eq
A^{\dagger}C = C^{\dagger}A\ , \qquad B^{\dagger}D = D^{\dagger}B\ , \qquad
A^{\dagger}D - C^{\dagger}B = 1\ ,
\feq
i.e., the complex $8\times8$ matrix
\eq
\left(\begin{array}{cc} A & B \\ C & D\end{array}\right) \in \text{Sp}(4,\bC)\ ,
\feq
and the sequence \eqref{sequence} is
\eq
\text{SU}(4) \subset \text{SL}(4,\bC) \subset \text{Sp}(4,\bC)\ .
\feq
The relevance of the M\"obius group lies in its interpretation as invariance group
of the BPS condition $\det Z=0$ \cite{Gauntlett:2000ch}: For the standard $D=4$,
$N=1$ Poincar\'e superalgebra without central charges, all BPS states obey
$P^2=0$. This is the momentum space version of the massless wave equation,
which is invariant under the conformal group $\text{SU}(2,2)$ in $3+1$ dimensions.
The identification of four-dimensional Minkowski space with the Jordan algebra
of hermitian $2\times2$ matrices leads then to an identification of this conformal group
with the M\"obius group of $J_2^{\bC}$. As a generalization of this, we have thus
found that the BPS condition for the AdS superalgebra \eqref{QiQj}, \eqref{superalg} is
preserved by the symplectic group $\text{Sp}(4,\bC)$.

In what follows, we would like to give a geometrical interpretation of the various subsets of
the cone containing states that preserve different fractions of supersymmetry.
To this end, we first note that, as in \cite{Gauntlett:2000ch},  the cone is a stratified space with
strata ${\cal S}_n$, $n=0,1,2,3,4$, where ${\cal S}_n$ denotes the subspace in which at least
$n$ of the four eigenvalues vanish, corresponding to at least $n$ supersymmetries being
preserved, and ${\cal S}_{n+1}$ is the boundary of ${\cal S}_n$. ${\cal S}_0$ is the convex
cone itself, consisting of all positive hermitian $4\times4$ matrices. The boundary of the cone
is the subspace ${\cal S}_1$, which is the 15-dimensional space of matrices of rank 3 or less.
The boundary of this, containing the states preserving at least one half of the supersymmetries,
is the space ${\cal S}_2$ of matrices of rank 2 or less, which has dimension 14. In order to see this,
recall that a matrix of rank two is completely specified by its two normalized eigenvectors belonging to
the nonvanishing eigenvalues, plus the non-zero eigenvalues themselves.
The two eigenvectors span a two-plane in $\bC^4$, i.e., they correspond to the Stiefel
manifold\footnote{The Stiefel manifold $V_k(\bb{K}^n)$ is the set of all orthonormal $k$-frames
in $\bb{K}^n$.} $V_2(\bC^4)$. Since $V_k(\bC^n)\approx\text{U}(n)/\text{U}(n-k)$, one has
$\text{dim}(V_k(\bC^n))=2nk-k^2$ and thus $\text{dim}(V_2(\bC^4))=12$.
Taking into account also the two eigenvalues, we get
\begin{displaymath}
{\cal S}_2 \approx (\text{U}(4)/\text{U}(2))\oplus (\bR^+)^2\ ,
\end{displaymath}
that has dimension 14, as claimed above. The boundary of ${\cal S}_2$ is the set ${\cal S}_3$
of matrices having rank one or less. These span a 7-dimensional space, since a rank one
matrix is specified by the direction, up to a sign, of its eigenvector with non-zero eigenvalue
together with the eigenvalue. This is a point in $\bC P^3\times\bR^+$.

\section{Final remarks}
\label{fin-rem}

We conclude this work with some comments and possible future developments.
First of all, it would be interesting to understand how the proposed superalgebra
su$(2,2|1)$ fits into $N=2$ gauged supergravity, whose supersymmetric solutions were
studied as examples. Our results might in fact point towards a hidden superconformal
invariance of this supergravity theory. In this context it is amusing to note that the
superconformal group plays a role in the construction of $N=2$, $D=4$ (gauged) supergravity
(cf.~\cite{Vambroes} for a review). This goes under the name of superconformal tensor calculus.

A second point is how our proposal could be generalized to the case with more vector
fields and, correspondingly, more magnetic charges, as it happens e.g.~in $N=2$ gauged
supergravity coupled to vector multiplets. Black holes with more than one magnetic charge
do indeed exist in these theories \cite{Cacciatori:2009iz}. The most obvious thing to do
would be to add a five-dimensional vector for each charge, but it remains to be seen if such
a proposal fits into some adequate superalgebra. A related topic is the inclusion of
magnetic charges in osp$(4|N)$ for $N>2$, which should lead to extended superconformal
algebras. A further investigation of these points will be presented elsewhere.

\acknowledgments

This work was partially supported by INFN and MIUR-PRIN contract 20075ATT78.
We would like to thank Antoine Van Proeyen for clarifying correspondence.

\appendix

\section{Conventions}
\label{conv}

Throughout this work, indices $i,j,\ldots$ range from 1 to 2, and $\epsilon_{ij}=-\epsilon_{ji}$, with
$\epsilon_{12}=1$. Capital latin indices $A,B,\ldots=0,\ldots,4$ refer to SO$(2,3)$ tensors.
The gamma matrices $\Gamma_{AB}$ satisfy
\eq
\{\Gamma_A,\Gamma_B\} = 2\eta_{AB}\ , \label{5dCliff}
\feq
where $\eta_{AB}=\text{diag}(-1,1,1,1,-1)$. For the Dirac matrices $\gamma^a$
($a=0,\ldots,3$) in four dimensions we choose the real representation
\begin{displaymath}
\gamma^0 = \left(\begin{array}{cc} 0 & i\sigma^2 \\ i\sigma^2 & 0\end{array}\right)\ , \qquad
\gamma^1 = \left(\begin{array}{cc} -\sigma^3 & 0 \\ 0 & -\sigma^3\end{array}\right)\ , \qquad
\gamma^2 = \left(\begin{array}{cc} 0 & -i\sigma^2 \\ i\sigma^2 & 0\end{array}\right)\ ,
\end{displaymath}
\begin{displaymath}
\gamma^3 = \left(\begin{array}{cc} \sigma^1 & 0 \\ 0 & \sigma^1\end{array}\right)\ , \qquad
\gamma^5 = \gamma^{0123} = \left(\begin{array}{cc} -i\sigma^2 & 0 \\ 0 & i\sigma^2
\end{array}\right)\ ,
\end{displaymath}
where $\sigma^i$ denote the standard Pauli matrices. Starting from this, one can construct
a realization of the SO$(2,3)$ Clifford algebra \eqref{5dCliff} by setting
\eq
\Gamma^A = \left\{\begin{array}{c@{\,, \quad}l} \gamma^5\gamma^a & A=a=0,\ldots,3\ , \\
                        \gamma^5 & A=4\ .\end{array}\right. \label{Gamma}
\feq
This implies
\eq
\Gamma^{AB} \equiv \frac12[\Gamma^A,\Gamma^B] = \left\{\begin{array}{c@{\,, \quad}l}
\gamma^{ab} & A=a\,, \, B=b\ , \\ \gamma^a & A=a\,, \, B=4\ .\end{array}\right.
\feq
The charge conjugation matrix $C$ in four dimensions satisfies $C^T=C^{-1}=-C$ and
\eq
{\gamma^a}^T = -C\gamma^aC^{-1}\ .
\feq
Using the definition \eqref{Gamma}, one shows that $C$ is then a charge conjugation
matrix in five dimensions as well, but with a change of sign,
\eq
{\Gamma^A}^T = C\Gamma^AC^{-1}\ .
\feq
Here we choose $C=\gamma^0$. Then the Majorana condition $\bar\epsilon=\epsilon^TC$,
where $\bar\epsilon=\epsilon^{\dagger}\gamma^0$, implies that the spinor $\epsilon$ is real.
Notice that the Majorana condition is preserved under SO$(2,3)$ transformations, since
\eqref{Gamma}, together with
\eq
{\gamma^a}^{\star} = B\gamma^a B^{-1}\ ,
\feq
where $B=-C\gamma^0$, implies
\eq
{\Gamma^{AB}}^{\star} = B\Gamma^{AB}B^{-1}\ .
\feq

\section{The superalgebra osp$(4|N)$}
\label{app-osp}

The superalgebra osp$(4|N)$ is defined as the set of graded $(4+N)\times(4+N)$ matrices
$\mu$ satisfying the conditions
\begin{eqnarray}
\mu^T\left(\begin{array}{cc} C & 0 \\ 0 & 1_{N\times N}\end{array}\right) &+&
\left(\begin{array}{cc} C & 0 \\ 0 & 1_{N\times N}\end{array}\right)\mu = 0\ , \nonumber \\
\mu^{\dagger}\left(\begin{array}{cc} \gamma^0 & 0 \\ 0 & -1_{N\times N}\end{array}\right) &+&
\left(\begin{array}{cc} \gamma^0 & 0 \\ 0 & -1_{N\times N}\end{array}\right)\mu = 0\ . \nonumber
\end{eqnarray}
These equations can be solved by setting
\eq
\mu = \left(\begin{array}{cc} \frac14\epsilon^{AB}[\Gamma_A,\Gamma_B] & \chi^i \\
\bar\chi^i & i\epsilon_{ij}t^{ij}\end{array}\right)\ ,
\feq
where $\epsilon^{AB}$ and $\epsilon_{ij}$ denote respectively an arbitrary real
antisymmetric 4$\times$4 and $N\times N$ tensor, $\Gamma_A$ are SO$(2,3)$ Dirac matrices,
$t^{ij}$ represent SO$(N)$ generators and $\chi^i$ is a set of $N$ Majorana spinors.
The bosonic subalgebra of osp$(4|N)$ is $\text{sp}(4)\!\oplus\text{so}(N)$ $\cong$
$\text{so}(2,3)\!\oplus\text{so}(N)$. In the case $N=2$ the so$(2)$ generator is interpreted
as electric charge.


\begin{thebibliography}{99}

\bibitem{Caldarelli:1998hg}
  M.~M.~Caldarelli and D.~Klemm,
  ``Supersymmetry of anti-de~Sitter black holes,''
  Nucl.\ Phys.\  B {\bf 545} (1999) 434
  [arXiv:hep-th/9808097].

\bibitem{Cacciatori:2009iz}
  S.~L.~Cacciatori and D.~Klemm,
  ``Supersymmetric AdS$_4$ black holes and attractors,''
  arXiv:0911.4926 [hep-th].

\bibitem{Romans:1991nq}
  L.~J.~Romans,
  ``Supersymmetric, cold and lukewarm black holes in cosmological
  Einstein-Maxwell theory,''
  Nucl.\ Phys.\  B {\bf 383} (1992) 395
  [arXiv:hep-th/9203018].

\bibitem{Kostelecky:1995ei}
  V.~A.~Kosteleck\'y and M.~J.~Perry,
  ``Solitonic black holes in gauged $N=2$ supergravity,''
  Phys.\ Lett.\  B {\bf 371} (1996) 191
  [arXiv:hep-th/9512222].

\bibitem{Argurio:2008zt}
  R.~Argurio, F.~Dehouck and L.~Houart,
  ``Supersymmetry and Gravitational Duality,''
  Phys.\ Rev.\  D {\bf 79} (2009) 125001
  [arXiv:0810.4999 [hep-th]].

\bibitem{Hull:1997kt}
  C.~M.~Hull,
  ``Gravitational duality, branes and charges,''
  Nucl.\ Phys.\  B {\bf 509} (1998) 216
  [arXiv:hep-th/9705162].

\bibitem{AlonsoAlberca:2000cs}
  N.~Alonso-Alberca, P.~Meessen and T.~Ort\'{\i}n,
  ``Supersymmetry of topological Kerr-Newman-Taub-NUT-adS spacetimes,''
  Class.\ Quant.\ Grav.\  {\bf 17} (2000) 2783
  [arXiv:hep-th/0003071].

\bibitem{West:1998ey}
  P.~C.~West,
  ``Supergravity, brane dynamics and string duality,''
  arXiv:hep-th/9811101.

\bibitem{Gauntlett:1999dt}
  J.~P.~Gauntlett and C.~M.~Hull,
  ``BPS states with extra supersymmetry,''
  JHEP {\bf 0001} (2000) 004
  [arXiv:hep-th/9909098].

\bibitem{Ferrara:1999si}
  S.~Ferrara and M.~Porrati,
  ``AdS$_5$ superalgebras with brane charges,''
  Phys.\ Lett.\  B {\bf 458} (1999) 43
  [arXiv:hep-th/9903241].

\bibitem{Lee:2004jx}
  S.~Lee and J.~H.~Park,
  ``Noncentral extension of the AdS$_5$ $\times$ S$^5$ superalgebra:
  supermultiplet of brane charges,''
  JHEP {\bf 0406} (2004) 038
  [arXiv:hep-th/0404051].

\bibitem{Gauntlett:2000ch}
  J.~P.~Gauntlett, G.~W.~Gibbons, C.~M.~Hull and P.~K.~Townsend,
  ``BPS states of $D=4$ $N=1$ supersymmetry,''
  Commun.\ Math.\ Phys.\  {\bf 216} (2001) 431
  [arXiv:hep-th/0001024].

\bibitem{Caldarelli:1999xj}
  M.~M.~Caldarelli, G.~Cognola and D.~Klemm,
  ``Thermodynamics of Kerr-Newman-AdS black holes and conformal field
  theories,''
  Class.\ Quant.\ Grav.\  {\bf 17} (2000) 399
  [arXiv:hep-th/9908022].

\bibitem{Vanzo:1997gw}
  L.~Vanzo,
  ``Black holes with unusual topology,''
  Phys.\ Rev.\  D {\bf 56} (1997) 6475
  [arXiv:gr-qc/9705004].

\bibitem{Emparan:1999gf}
  R.~Emparan,
  ``AdS/CFT duals of topological black holes and the entropy of  zero-energy
  states,''
  JHEP {\bf 9906} (1999) 036
  [arXiv:hep-th/9906040].

\bibitem{Balasubramanian:1999re}
  V.~Balasubramanian and P.~Kraus,
  ``A stress tensor for anti-de~Sitter gravity,''
  Commun.\ Math.\ Phys.\  {\bf 208} (1999) 413
  [arXiv:hep-th/9902121].

\bibitem{Gunaydin:1992zh}
  M.~G\"unaydin,
  ``Generalized conformal and superconformal group actions and Jordan
  algebras,''
  Mod.\ Phys.\ Lett.\  A {\bf 8} (1993) 1407
  [arXiv:hep-th/9301050].

\bibitem{Vambroes}
 A.~Van Proeyen,
 ``${\cal N}=2$ supergravity in $d=4,5,6$ and its matter couplings,''
 extended version of lectures given during the semester ``Supergravity,
 superstrings and M-theory'' at Institut Henri Poincar\'e, Paris, november 2000;
 http://itf.fys.kuleuven.ac.be/$\sim$toine/home.htm\#B



\end{thebibliography}
\end{document}